\documentclass[pra,aps,groupaddress,twocolumn,floatfix]{revtex4}
\usepackage{amsmath,amsfonts,amssymb,graphics,graphicx,epsfig,color,times,natbib}

\newcommand{\be}{\begin{equation}}
\newcommand{\ee}{\end{equation}}
\newcommand{\ba}{\begin{eqnarray}}
\newcommand{\ea}{\end{eqnarray}}
\newcommand{\ban}{\begin{eqnarray*}}
\newcommand{\ean}{\end{eqnarray*}}

\newcommand{\ket}[1]{\mbox{$ | #1 \rangle $}}
\newcommand{\bra}[1]{\mbox{$ \langle #1 | $}}

\newcommand{\si}{\sigma}

\newcommand{\one}{\leavevmode\hbox{\small1\normalsize\kern-.33em1}}

\begin{document}

\title{Can one see entanglement ?}
\author{Nicolas Brunner}
\email{n.brunner@bristol.ac.uk}
\author{Cyril Branciard}
\author{Nicolas Gisin}
\address{Group of Applied Physics, University of Geneva, CH-1211 Geneva 4, Switzerland}
\date{\today}

\begin{abstract}
The human eye can detect optical signals containing only a few
photons. We investigate the possibility to demonstrate
entanglement with such biological detectors. While one person could
not detect entanglement by simply observing photons, we discuss
the possibility for several observers to demonstrate entanglement
in a Bell-type experiment, in which standard detectors are
replaced by human eyes. Using a toy model for biological detectors
that captures their main characteristic, namely a detection
threshold, we show that Bell inequalities can be violated, thus
demonstrating entanglement. Remarkably, when the response function
of the detector is close to a step function, quantum non-locality
can be demonstrated without any further assumptions. For smoother
response functions, as for the human eye, post-selection is
required.
\end{abstract}
\maketitle

\section{Introduction}

The human eye is an extraordinary light sensitive detector. It can
easily stand a comparison to today's best man-made detectors \cite{rieke}.
Already back in the forties, experiments on the sensibility of the
human eye to weak optical signals were conducted
\cite{pirenne}, leading to the conclusion that rod photoreceptors
can detect a very small number of photons, typically less than 10
during an integration time of about 300ms \cite{Barlow56}. To date, this
prediction has been confirmed by many experiments \cite{rieke}.
Though most specialists still disagree on the exact number of
photons required to trigger a neural response, it seems to be now
commonly accepted that there is a threshold number of incident
photons, below which no neural signal is sent to the brain. This
assumption is supported by the good agreement between theoretical
models and experimental data from behavioral experiments. Our visual system works basically as follows: first a photon is absorbed by the rod, which then amplifies the signal with some very efficient chemical reactions;
then some post-processing (basically a thresholding) is performed on the signals incoming from a group of 20-100 rods \cite{ThresholdReview}; finally a neural signal is eventually sent to the brain. The role of the threshold is possibly to maintain a very low dark
noise in the visual process, in particular to get rid of electrical noise originating from
the individual rods \cite{ThresholdReview,Threshold}.

\begin{figure}[b!]
\includegraphics[width=0.9\columnwidth]{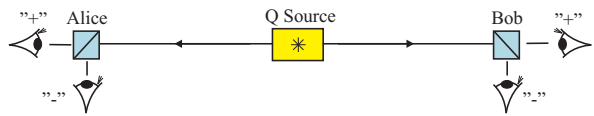} \caption{(Color online) Bell experiments with human detectors.}\label{seeEnt}
\end{figure}

In Quantum Information, experiments carried out on photons are now
routinely performed for demonstrating fascinating quantum
features, such as entanglement and quantum non-locality
\cite{Aspect99,ReviewPhotonic}. In this context, and considering the
amazing performances of the human eye, it is quite intriguing to
ask whether one could demonstrate entanglement without the help of
man-made detectors, but using only naked eyes. It should be
reminded at this point that entanglement is usually demonstrated
in Bell-type experiments (where correlations between two distant
parties are measured), and not via single shot measurements.
Therefore one person cannot expect to see entanglement directly.
Nevertheless, one could perform a Bell experiment in which
man-made photon detectors are replaced by human eyes (see Fig. 1),
or more generally by biological detectors. In case the collected
data would lead to the violation of a Bell's inequality
\cite{Bell64}, one could argue that entanglement has been "seen".
Let us stress that, though such an experiment would probably
not lead to a better understanding of quantum non-locality itself,
it would definitely be fascinating !

The main difference between man-made photon counters and the human
eye is a detection threshold. To test whether a detector is able
to detect single photons, one usually checks that the response of
this detector to very low intensities is linear. Indeed this is
not the case for the eye, where the efficiency of detection plotted as a function of the number of incoming photons is a typical S-shaped curve (see \cite{rieke}). In this paper we report a preliminary
theoretical study of Bell tests with threshold detectors. Our goal
is to provide a good understanding of Bell experiments with a toy
model for the detector that captures the main characteristic of
the human eye.

The presentation is organized as follows. After some general description of the scenario we consider (Section II), we first focus on
detectors with a perfect threshold, i.e. no detection below the threshold and perfect
detection above (Section III). We show that, even for a poissonian source, the
threshold is not a restriction for demonstrating quantum
non-locality; in other words, Bell inequalities can be violated
(in the strict sense) with such detectors. Then we smoothen the
threshold, in order to make our detector model closer to the human eye (Section IV). We show
that, except for close to perfect thresholds, one must then
perform post-selection in order to obtain a violation of a Bell inequality (Section V).

\section{General framework}

Let us consider a typical Bell test
scenario. A source sends pairs of entangled particles (each pair
being in state $\rho$) to two distant observers, Alice and Bob,
who perform measurements on their respective particles. Here,
Alice and Bob choose between two different measurement settings
$A_{1}$, $A_{2}$ and $B_{1}$, $B_{2}$, each of these measurements
giving a binary result $\alpha,\beta \in \{+,-\}$. In this case
the relevant Bell inequality is the famous Clauser-Horne-Shimony-Holt (CHSH)
inequality \cite{chsh}, which we will express here in the Clauser-Horne (CH)
\cite{ch} form \ba\label{CH} \nonumber CH &\equiv&
P_{++}(A_{1}B_{1})+P_{++}(A_{1}B_{2})+P_{++}(A_{2}B_{1})
\\  & & -P_{++}(A_{2}B_{2})-P_{+}(A_{1})-P_{+}(B_{1})  \leq 0 \,\, , \ea
where $P_{++}(A_{i}B_{j}) \equiv P(++|A_{i}B_{j})$ is
the probability that $\alpha=\beta=+$ when Alice (Bob) has
performed measurement $A_i$ ($B_j$). Note that under the
hypothesis of no-signaling, inequalities CH and CHSH are strictly
equivalent \cite{dan}.

\section{Perfect threshold}

Now let us bring the threshold detector
into the picture. We start by considering a detector with a
perfect threshold at $N$ photons; optical signals containing at
least $N$ photons are always detected (note that our detector is
not photon number resolving), while signals with less than $N$
photons are never detected. The response function of our detector
is simply a step function, the step occurring at $N$ photons.

First, it is clear that the number of emitted pairs $M$ has to be
larger or equal than the threshold $N$, otherwise the detectors
will never fire. At this point it should be reminded that Bell inequalities are usually considered in a situation where the source emits a single pair of entangled particles at a time. Nevertheless, the violation of Bell inequalities can also be studied in the multi-pair scenario \cite{drummond83,reid02,Multipair}; of particular interest are experimental situations where single entangled pairs cannot be individually created or measured, for instance in many-body systems \cite{Multipair}. Nevertheless such studies require a careful analysis, in particular when post-selection is performed, as we shall see in Section V.

To gain some intuition, let us start with the
simplest situation $M=N$: the source emits exactly the threshold
number of pairs. In this case a detector clicks whenever all
photons take the same output of the polarizing beam splitter. Thus
the probabilities entering the CH inequality are simply given by
\ba\label{probs} P_+(A_i)= p_+(A_i)^N \,\, , \,\, P_{++}(A_iB_j)=
p_{++}(A_iB_j)^N \,\, , \ea where $p_{++}(A_iB_j)=tr( [A_i^+
\otimes B_j^+ ] \rho)$ is the quantum joint probability for a
single pair to give a click in the "+" detector on Alice's and on
Bob's side, and similarly for the marginal probability
$p_{+}(A_i)$. It should be stressed that, though the detectors are
supposed to be perfectly efficient, there are many inconclusive
events "$\emptyset$" (not giving any click in detector "+" or in
detector "-"), because of the threshold. Since we consider only
the outcome "+" in the CH inequality, one may relabel the outcomes
in the following way: "+" $\rightarrow$ "+" and "-,$\emptyset$"
$\rightarrow$ "0". Then, the experiment still provides binary
outcomes, "+" or "0", but no events have been discarded.

\begin{figure}
\includegraphics[width=0.9\columnwidth]{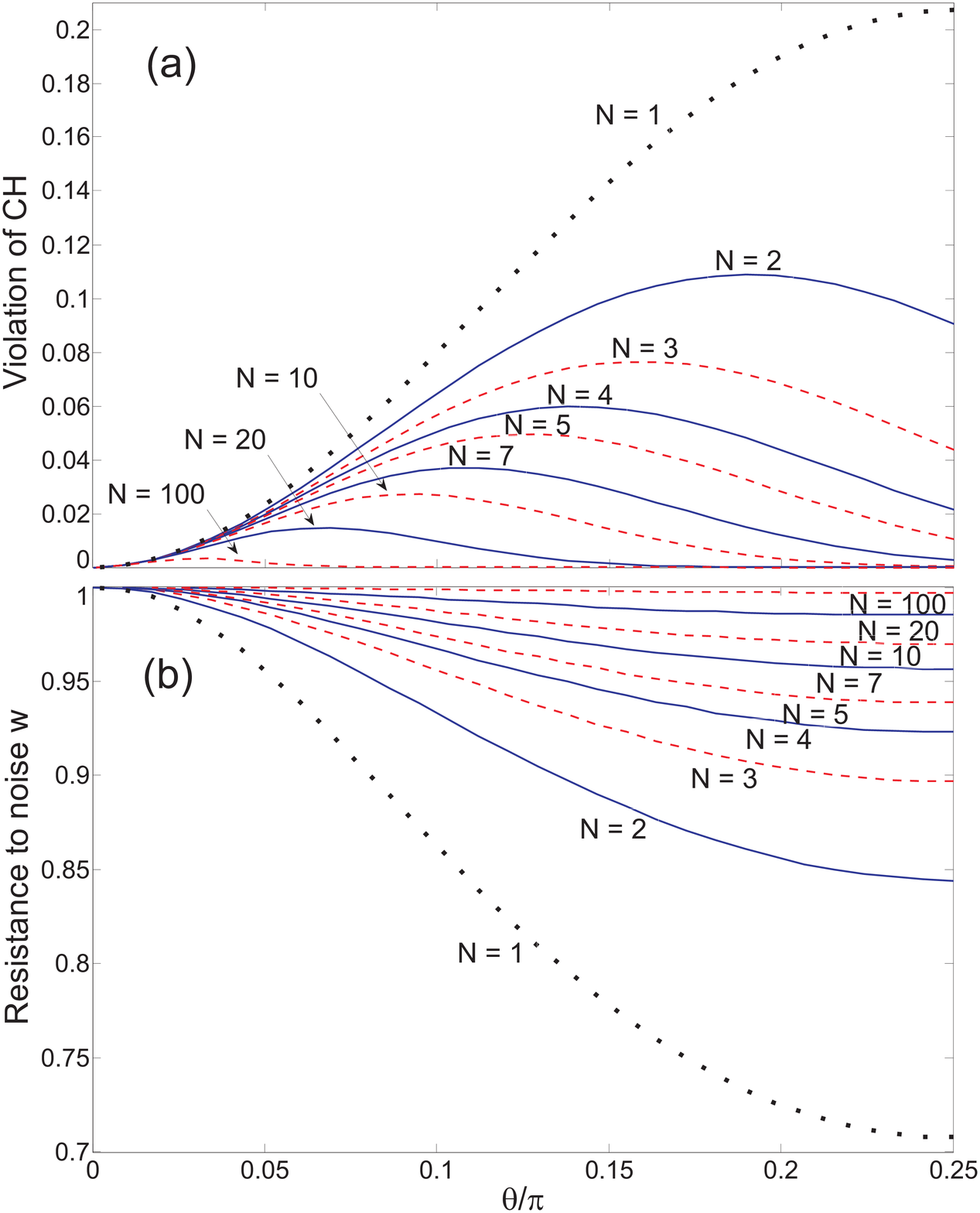} \caption{(Color online) Violation of the CH
inequality (a) and resistance to noise $w$ (b) versus the degree
of entanglement of the state, for different threshold values $N$.
Remarkably the inequality is violated for any value of the
threshold $N$.}\label{CH_N}
\end{figure}

Inserting probabilities (\ref{probs}) into the CH inequality, we
compute numerically the maximal amount of violation for pure
entangled states of two qubits $\ket{\psi}=\cos{\theta}\ket{00} +
\sin{\theta}\ket{11}$. The optimization is performed over the four
measurement settings. The results are presented in Fig. \ref{CH_N}
(a), for different values of the threshold $N$. Surprisingly, the
inequality can be violated for any $N$; this can be shown analytically for the maximally entangled state, see \footnote{Consider for example the maximally
entangled state, and the settings $A_1 = \sigma_z$, $A_2 =
\cos(2\varphi)\sigma_z+\sin(2\varphi)\sigma_x, B_1 =
\cos(\varphi)\sigma_z+\sin(\varphi)\sigma_x, B_2 =
\cos(\varphi)\sigma_z-\sin(\varphi)\sigma_x$. One gets $CH \sim
\frac{3N}{2^{N+1}} \varphi^2 >0$ when $\varphi \to 0$. Thus the CH inequality can be
violated for all $N$, with $\varphi$ small enough.}. For large
values of $N$, this is quite astonishing, since the probabilities
(\ref{probs}) are very small; most events do not lead to a click
in the "+" detector. Let us stress that no particular assumptions
(such as fair-sampling) are required here, since no events have
been discarded. Another astonishing feature, is that, for
increasing values of $N$, the state that achieves the largest
violation is less and less entangled. Note that in general,
the relation between entanglement and non-locality is not well
understood, but hints suggest that partially entangled state
contain more non-locality than maximally entangled ones
 \cite{methotscarani,CabelloDetLoop,AsymDetLoop}.

Next we compute the resistance to noise, defined as the maximal
amount ($1-w$) of white noise that can be added to the state
$\ket{\psi}$ such that the global state $\rho = w
\ket{\psi}\bra{\psi}+ (1-w)\frac{\one}{4}$ still violates the Bell
inequality. The optimization is performed as above. We find that
the more entangled the state is, the more robust it is, though for
$N\geq 2$ the maximal violation is not obtained for the maximally
entangled state (see Fig. \ref{CH_N}). So the close relation that
exists, in the standard case $N=1$, between the amount of
violation and the resistance to noise, does not hold here anymore
\footnote{In the case $N=1$ the resistance to noise $w$ can be
expressed as a function of the amount of violation $Q$:
$w=\frac{L-\mathcal{M}}{Q-\mathcal{M}}$, where
$Q=tr(\mathcal{B}\ket{\psi}\bra{\psi})$,
$\mathcal{M}=tr(\mathcal{B})/4$ and $\mathcal{B}$ is the Bell
operator, $L$ the local bound of the inequality.}. Indeed, in the
perspective of experiments, the resistance to noise is the
relevant figure of merit.

Now let us consider the case where the source emits $M \geq N$
entangled pairs, and the detector is characterized by a response
function $\Theta(x)$, where $x$ is the number of incident photons.
The probabilities (\ref{probs}) now read \ba\label{probsM}
P_{+}^{(M)}&=&\sum_{n_{+}+n_{-}=M} \Theta(n_+) \, M! \,
\frac{p_+^{n_+}}{n_+!} \, \frac{p_-^{n_-}}{n_-!}
\\\nonumber P_{++}^{(M)}&=&\sum_{\sum{n_{\alpha\beta}=M}}
\Theta(n_{+}^A)\, \Theta(n_{+}^B)\, M! \prod_{\alpha,\beta=\pm}
\left(
\frac{p_{\alpha\beta}^{n_{\alpha\beta}}}{n_{\alpha\beta}!}\right)
\ea where the indices $n_{\alpha,\beta}$ represent the numbers of
pairs that take the outputs $\alpha$ on Alice's side and $\beta$
on Bob's side, and $n_{+}^A \equiv n_{++} + n_{+-}$ while $n_{+}^B
\equiv n_{++} + n_{-+} $.

\begin{figure}
\includegraphics[width=0.9\columnwidth]{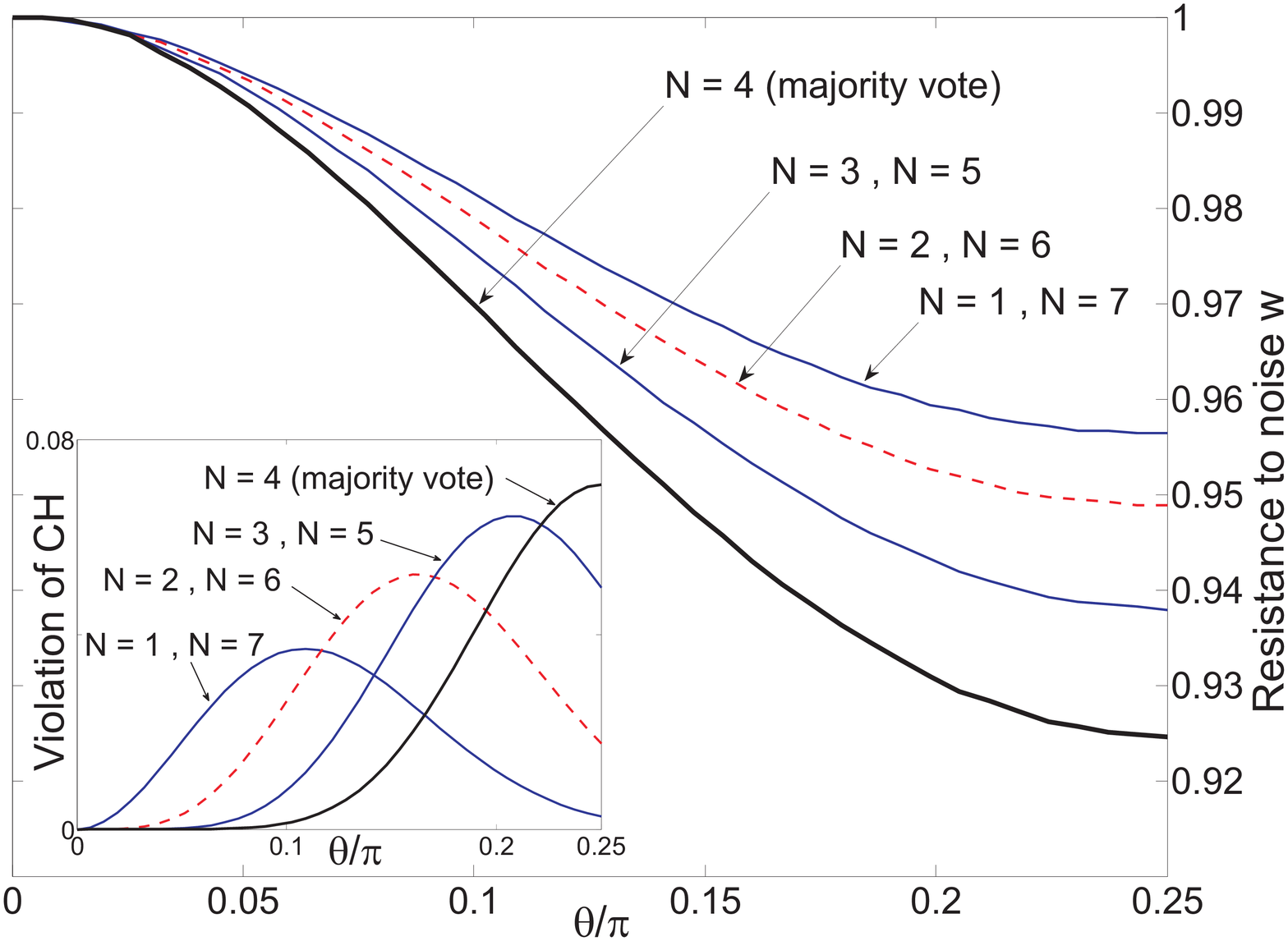} \caption{(Color online) Violation of the CH inequality (inset)
and the resistance to noise $w$ (figure) versus the degree of
entanglement of the state, for different thresholds $N$. The
source emits a fixed number of pairs $M=7$. The optimal threshold,
$N=4$, corresponds to a majority vote (see text).}\label{CH_M}
\end{figure}

For now, we still consider detectors with a perfect threshold,
i.e. $\Theta(x<N)=0$ and $\Theta(x\geq N)=1$. Again, the amount of
violation of the CH inequality as well as the resistance to noise
can be computed numerically. We have performed optimization for
$N\leq10$ and found that the CH inequality can still be violated
but that the resistance to noise decreases for increasing values
of $M$. In Fig. \ref{CH_M} we present the results in a slightly
different way: for a fixed number of emitted pairs ($M=7$), we
compute the violation of CH and the resistance to noise for
different thresholds $N$. The optimal threshold is found to be
$N=\lfloor\frac{M+1}{2}\rfloor$. Note that if we had photon
counting detectors, then this threshold would simply correspond to
a majority vote \cite{Multipair}: if $n_+ \geq n_-$ then the result is "+",
otherwise it is "-". It should also be pointed out that detectors
with threshold $N$ and $M-N+1$ are equivalent, which can be seen
by inverting the outputs "+" and "-" \footnote{For a
threshold $M-N+1$, having a click in detector "-" implies that
$n_{-} \geq M-N+1$. We suppose that in case both detectors fire,
the result "+" is outputted. Thus, to get output "-", one must add
the constraint that $n_+ <M-N+1$, which implies that $n_- \geq N$,
since $n_{+} + n_{-} = M$. So models with thresholds $N$ and
$M-N+1$ are made equivalent, by inverting the outputs "+" and "-".
This also shows that the double click events play no role. If in
case of a double click, the outcome "+" is given, then considering
the output "-" provides a model without the double click.}.

\begin{figure}[t!]
\includegraphics[width=0.9\columnwidth]{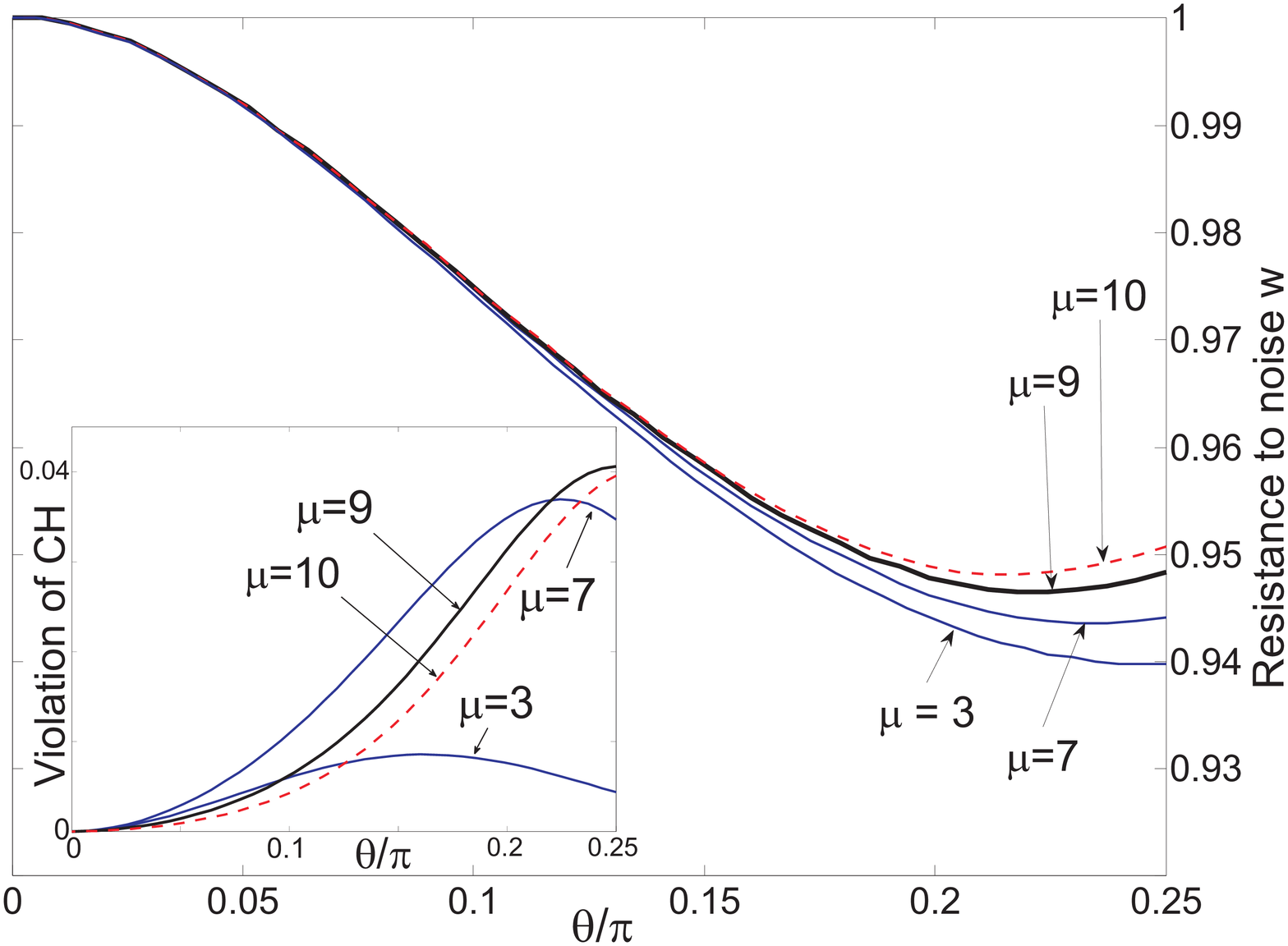} \caption{(Color online) Violation of CH (inset)
and resistance to noise $w$ (figure) versus the degree of
entanglement of the state, for a poissonian source. The threshold
is fixed to $N=5$, while the mean number of emitted pairs $\mu$ is
varied.}\label{CH_poisson}
\end{figure}

Next we consider a poissonian source. The probability of emitting
$M$ pairs is $p_{M}=e^{-\mu}\frac{\mu^M}{M!}$, where $\mu$ is the
mean number of emitted pairs. Again we compute the probabilities
entering the CH inequality: \ba\nonumber P_{+}^{(\mu)} &=& \sum_{M} p_{M}
P_{+}^{(M)} \\ P_{++}^{(\mu)} &=& \sum_{M} p_{M} P_{++}^{(M)} \,\, , \ea with
$P_{+}^{(M)}$ and $P_{++}^{(M)}$ defined in equations
(\ref{probsM}). Numerical optimizations show that the CH
inequality can be violated. Fig. \ref{CH_poisson} shows the
results for a detector with a perfect threshold at $N=5$. The
largest violation is obtained for $\mu \approx 9.05 \approx 2N-1$,
so basically when the threshold corresponds to a majority vote on
the mean number of pairs ($N \approx \frac{\mu+1}{2}$). The
resistance to noise has a very different dependance on $\mu$ (see
Fig. \ref{CH_poisson}). Smaller values of $\mu$ are more robust
against noise. Intuitively this can be understood as follows. The
term with $M=N$ pairs is the most robust against noise, as
discussed previously. For small values of $\mu$, more weight is
given to this term (compared to terms with more pairs), thus
leading to a stronger resistance to noise.

\section{Smooth threshold}

We just showed that a threshold is not
a limiting factor for demonstrating quantum non-locality, and consequently entanglement. However
the response function of real biological detectors, such as the
human eye, is not a perfect threshold but a smooth curve.
Typically, for a number of photons near the threshold, the
efficiency is low; for instance $\sim 20\%$ for 60 incoming photons (see \cite{rieke} for details); here the
threshold being the minimum number of photons that can be detected
with a strictly positive probability. Let us also stress that the efficiency of the human eye does strongly depend on the number of incoming photons. Therefore the probability of seing cannot be characterized by a single parameter (for instance the efficiency for a single photon) as it is the case for linear detectors. One must consider the eye's (S-shaped) response function.

We have checked that, for smooth thresholds, the demonstration of quantum non-locality in
the strict sense is compromised, except if the response function
is close to a step function. Therefore, in the case of a response function with a smooth threshold, for instance in the case of the human eye, post-selection must be performed.

\section{Post-selection}

We post-select only the events leading to
a conclusive result on both sides, i.e. when one detector on
Alice's side and one detector on Bob's side fire. In this case
probabilities must be renormalized such that
$\bar{P}(\alpha\beta|ij)=P(\alpha\beta|A_iB_j)/\sum_{\alpha,\beta=
\pm}P(\alpha\beta|A_iB_j)$. Since we post-select coincidences, it
is now more convenient to express the CHSH inequality in its
standard form (in which only correlation terms appear) \ba S
&\equiv& |E(A_1,B_1)+E(A_1,B_2)
\\\nonumber & &+E(A_2,B_1)-E(A_2,B_2)| \leq L \, , \ea where $E(A_i,B_j)=
\sum_{\alpha,\beta=\pm } \alpha \beta \bar{P}(\alpha
\beta|A_iB_j)$, and $L$ is the local bound (for a single pair
$L=2$). Let us stress already that because of the post-selection,
the local bound $L$ for multi-pairs will be modified (see below).

The detector has now a smooth threshold at $N$ photons: below the
threshold the efficiency is zero $\Theta(x<N)=0$, at the threshold
the efficiency is limited $0<\Theta(x=N)<1$, and the efficiency
above the threshold is for now arbitrary. We start again with the
case where the source sends exactly $N$ pairs. The source is
supposed to send multiple copies of the same state $\rho$. For the
singlet state ($\rho = \ket{\psi^-}\bra{\psi^-}$), one has that
$p_{\psi_-}(\alpha \beta |\vec{a}\vec{b})=(1 - \alpha \beta
\vec{a} \cdot \vec{b} )/4$; here measurement settings are written as vectors on the Bloch sphere. This leads to \ba\label{correlPS}
E^{(N)}(\vec{a},\vec{b})=\frac{ (1 - \vec{a} \cdot \vec{b} )^N  -
(1 +  \vec{a} \cdot \vec{b} )^N}{(1 - \vec{a} \cdot \vec{b} )^N  +
(1 +  \vec{a} \cdot \vec{b} )^N} \,\, .\ea These correlations are
stronger than those of quantum physics for a single pair. This is
a consequence of the post-selection we performed. Note also, that
our post-selection depends (in general) on the measurement
settings, therefore the local bound $L$ of the CHSH inequality
must be modified accordingly. Inserting the correlators
(\ref{correlPS}) into the CHSH inequality, one gets an expression
which is maximized by the usual optimal settings, i.e.
$A_1=\si_z$, $A_2=\si_x$, $B_1=(\si_z+\si_x)/\sqrt{2}$ and
$B_2=(\si_z-\si_x)/\sqrt{2}$. In this case one gets

\ba\label{singlet} S^{(N)}_{\psi_-}=4 \frac{(1+1/\sqrt{2})^N -
(1-1/\sqrt{2})^N}{(1+1/\sqrt{2})^N + (1-1/\sqrt{2})^N} \, . \ea
For $N\geq 2$, $S^{(N)}_{\psi_-}$ exceeds the Tsirelson bound
($2\sqrt{2}$) \cite{tsirelson}: for example, $S^{(2)}_{\psi_-}
=8\sqrt{2}/3\approx 3.77$. In fact \eqref{singlet} tends to the algebraic limit of CHSH, $\lim_{N
\rightarrow \infty}S^{(N)}_{\psi_-}=4$.

\begin{figure}[t!]
\includegraphics[width=\columnwidth]{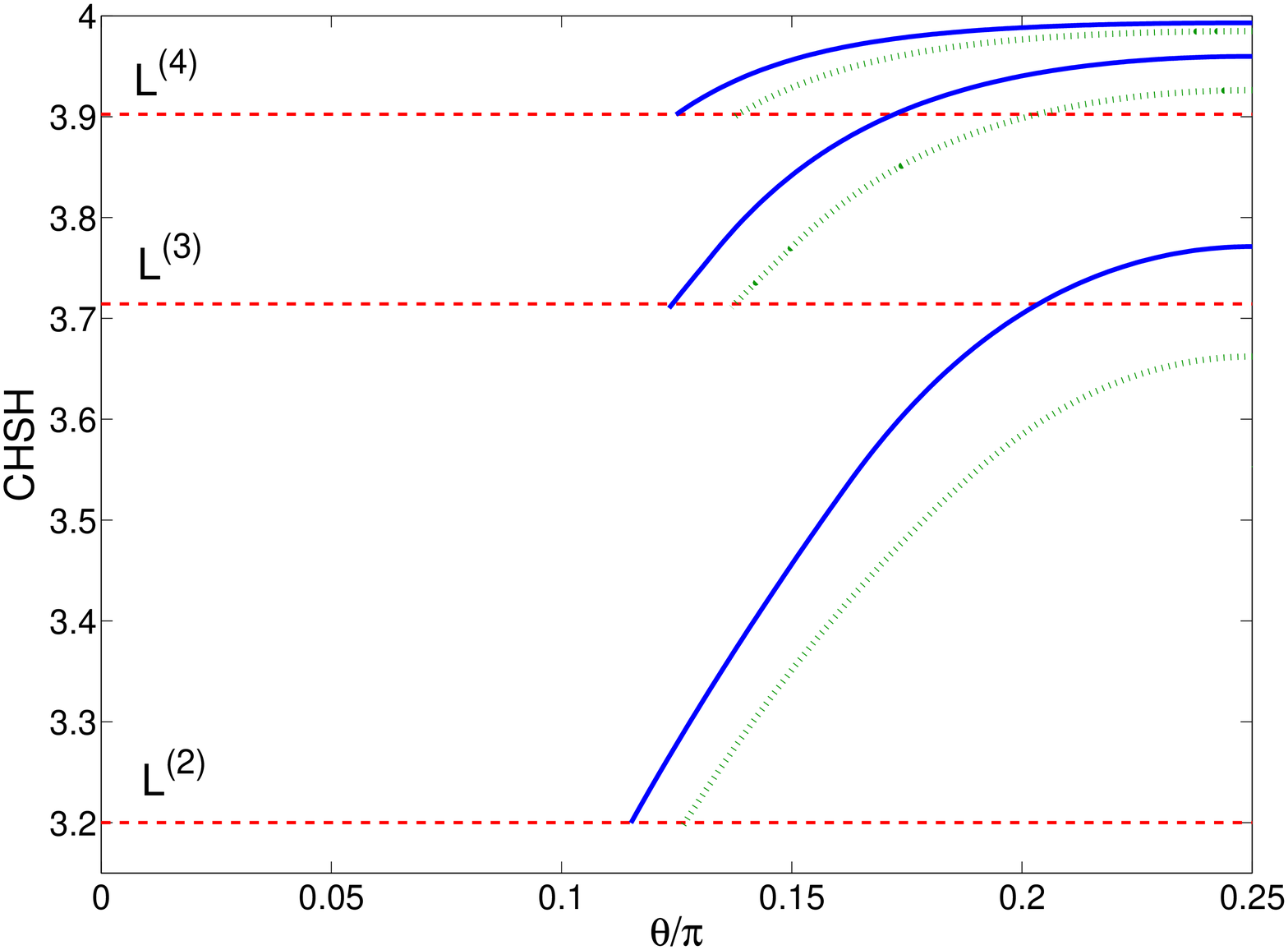}
\caption{(Color online) Violation of the CHSH inequality for different thresholds
$N$. The local bound is a function of $N$ (see text). Violations
are obtained for a source emitting exactly $N$ pairs (solid blue
lines) as well as for a poissonian source with $\mu=0.1$ (dotted
green lines). Note also that the set of pure entangled states that
violate CHSH becomes smaller for increasing $N$. The settings are
optimized for all states.}\label{CHSH_EYE}
\end{figure}

Thus we find that the violation of CHSH for the singlet state
increases with the threshold $N$. However, in order to conclude
for the presence of entanglement one has still to find the bound
for separable states. For $N=1$ this bound is indeed equal to the
local limit of the inequality (L=2). In case $N\geq 2$, the
local bound will be increased because of the post-selection, as
intuition suggests. Next we compute this local bound, which is
indeed also a bound for any separable state of the form
$\rho^{\otimes N}$.

We proceed as follows. We perform a numerical optimization over
any local probability distribution for two binary settings on each
side. This probability distribution is of the form of $N$ copies
of a (2-input/2-output) local probability distribution, because of
our hypothesis. The largest value of $S^{(N)}$ is obtained for the
following probability distribution \footnote{We conjecture that
\eqref{probsW} gives the largest violation of CHSH (we checked it
for small values of $N$), since it is obtained by an equal mixture
of the eight determistic strategies saturating CHSH; geometricaly,
it sits in the center of the CHSH facet.}:

\ba\label{probsW}\nonumber  p(\alpha=\beta|ij)=3/4  &,& p(\alpha
\neq \beta|ij)=1/4 \, , \textrm{ if  } i=1 \textrm{ or } j=1
\\  p(\alpha=\beta|ij)=1/4  &,&  p(\alpha \neq
\beta|ij)=3/4 \, ,  \textrm{ if  } i=j=2 \, , \ea leading to the
local bound \ba\label{LHV_N} L^{(N)}=4 \bigg[ \frac{3^N-1}{3^N+1}
\bigg] \quad .\ea One can check that $S^{(N)}_{\psi_-}>L^{(N)}$
for any $N$  (see Fig. \ref{CHSH_EYE}).

Remarkably, probability distribution (\ref{probsW}) is obtained
quantum mechanically by performing the optimal measurements
(mentioned above) on the Werner state \cite{werner} ($\rho_w =
w\ket{\psi_-}\bra{\psi_-}+ (1-w)\one/4$) for
$w=\frac{1}{\sqrt{2}}$, i.e. when $\rho_w$ ceases to violate the
CHSH inequality. Thus the resistance to noise for the singlet
state, is independent of $N$. It would be interesting to see if a strictly lower bound exists for separable states.

Note however that the bound
(\ref{LHV_N}) is valid only under the assumption that the source
sends multiple copies of the same state $\rho$. In case this
assumption breaks, the local bound reaches the algebraic limit of
CHSH (L=4), thus removing any hope of demonstrating
entanglement. More precisely, there is a local model giving $L=4$ for all
$N \geq 2$. Whether this bound can be reached by a separable two-qubit state
is unclear. We stress that this was not the case for perfect
thresholds; there no assumption had to be made on the source.

Curiously no violation is obtained when the source sends a fixed
number of pairs larger than the threshold ($M>N$) \footnote{Here
we have considered the following response function:
$\Theta(x<N)=0$, $\Theta(x=N)=\eta>0 $ and $\Theta(x>N)=1$.}.
However for a poissonian source, CHSH can be violated for small
values of $\mu$, the mean number of emitted pairs (see Fig.
\ref{CHSH_EYE}). Intuitively, if $\mu<<N$, the term with $N$ pairs
is dominant. When $\mu \rightarrow 0$ , the curve
$M=N$ is recovered. Note that for a poissonian source, the local bound
must be defined carefully, since the number of emitted pair
varies. However when $\mu<<N$ it is reasonable to consider the
local bound $L^{(N)}$.

\section{Conclusion}

Amazed by the performances of the human eye,
which can detect a few photons, we investigated whether biological
detectors might replace man-made detectors in Bell-type
experiments. We showed that the main characteristic of these
detectors, namely a detection threshold, is not a restriction for
violating Bell inequalities. In particular, we showed that closed to perfect threshold detectors can be used to test quantum non-locality without the need of any supplementary assumption, such as fair-sampling. For detectors with a smoother response function, one must perform post-selection, but Bell inequalities can still be violated, thus highlighting the presence of entanglement under reasonable assumptions. These results represent a first encouraging step, since there is apparently no fundamental restriction to detect entanglement with threshold detectors. Nevertheless, the next crucial step will be to estimate the feasibility of such
experiment with realistic parameters.

The authors thank J.D. Bancal, V. Scarani and C. Simon for
discussions. We aknowledge financial support from the EU project
QAP (IST-FET FP6-015848) and Swiss NCCR Quantum Photonics.

\bibliographystyle{prsty}
\bibliography{C:/BIB/thesis}

\end{document}